\documentclass[pre,aps,superscriptaddress,twocolumn]{revtex4}

\usepackage[dvips]{graphics}
\usepackage{graphicx}
\usepackage{amsfonts}
\usepackage{amssymb}
\usepackage{amsmath}
\usepackage{subfigure}

\begin{document}

\title{Vortex Structures Formed by the Interference of Sliced Condensates}
\author{R. Carretero-Gonz{\'a}lez}
\affiliation{
Nonlinear Dynamical Systems Group\footnote{URL: {\tt http://nlds.sdsu.edu/}},
Department of Mathematics and Statistics,
and Computational Science Research Center,
San Diego State University, San Diego CA, 92182-7720, USA}
\author{N. Whitaker}
\affiliation{Department of Mathematics and Statistics, University of
Massachusetts, Amherst MA 01003-4515}
\author{P. G.\ Kevrekidis}
\affiliation{Department of Mathematics and Statistics, University of
Massachusetts, Amherst MA 01003-4515}
\author{D.J. Frantzeskakis}
\affiliation{Department of Physics, University of Athens, Panepistimiopolis, Zografos,
Athens 15784, Greece }
\
\begin{abstract}
We study the formation of vortices, vortex necklaces and vortex ring
structures as a result of the interference of higher-dimensional 
Bose-Einstein condensates (BECs).
This study is motivated by earlier theoretical results pertaining to
the formation of dark solitons by interfering quasi one-dimensional BECs,
as well as recent experiments demonstrating the formation of vortices by
interfering higher-dimensional BECs. Here, we
demonstrate the genericity of the relevant scenario, but also
highlight a number of additional
possibilities emerging in higher-dimensional settings.
A relevant example is, e.g.,
the formation of a ``cage'' of vortex rings surrounding
the three-dimensional bulk of the condensed atoms.
The effects of the relative phases of the different BEC fragments and
the role of damping due to coupling with the thermal cloud
are also discussed. Our predictions should be
immediately tractable in currently existing experimental BEC setups.
\end{abstract}

\date{To appear in {\em Phys.~Rev.~A}}

\maketitle

\section{Introduction.}

Shortly after the realization of atomic Bose-Einstein condensates (BECs)
a remarkable experiment was reported \cite{science}, establishing BECs as coherent matter waves.
This experiment demonstrated (apart from self-interference) the interference between two BECs
confined in a trap, which was divided into two separate parts by means of a repulsive ``hump''-shaped potential
induced by a laser beam (usually called ``light sheet''). The BECs
were left to expand and overlap forming interference fringes, similar to the ones known in optics.
Analysis of this phenomenon
\cite{rhorl} allowed for a quantitative
understanding of some of its key features (such as the fringe separation) based on the
mean-field theory \cite{review}, and in particular
the time-dependent Gross-Pitaevskii (GP) equation; the latter,
can be expressed in the following
dimensionless form (see, e.g., \cite{mplb}),
%
\begin{eqnarray}
i \frac{\partial u}{\partial t} = -\frac{1}{2} \nabla^2 u + V({\bf r}) u + |u|^2 u,
\label{ceq1}
\end{eqnarray}
where $u({\bf r},t)$ is the
condensate wavefunction (and $|u({\bf r},t)|^2$
is the atom density),
while $V({\bf r})$
is the trapping potential.
In fact, this potential is
also
time-dependent since
it has initially a double-well form
(and the condensate is allowed to relax to its ground state)
and subsequently, at $t=0$, the ``hump'' separating the two wells
is lifted;
this way, the two fractions of the BEC
are allowed to interfere
and produce the beautiful experimental pictures observed
(see, e.g., the observed pattern in Fig.~2 of Ref.~\cite{science}).
Such experiments and relevant theoretical studies are of particular value in this setting as they allow
the study of quantum phenomena
at the mesoscopic scale, but with an interesting additional twist:
while the underlying quantum
processes are purely linear,
in the mean-field
picture, inter-atomic interactions are accounted for through an effective
nonlinearity [see the last term in Eq.~(\ref{ceq1})]
that significantly enriches the linear behavior.

An important modification of the linear behavior introduced by the above mentioned
nonlinearity is that the underlying nonlinear GP model supports
``fundamental'' nonlinear structures in the form of
matter-wave solitons and vortices. Importantly, such structures, and particularly
dark \cite{dark}, bright \cite{bright} and gap solitons \cite{markus},
as well as vortices \cite{vortex} and vortex lattices \cite{vl}
have been observed in a series of experiments over the past decade.
Moreover, experimental observation
of dynamical features of these structures, such as
the decay of
dark solitons into vortex structures
\cite{DS_to_vortex} has also been reported.

Here, we will focus on BECs with repulsive inter-atomic interactions,
which support stable dark solitons and vortex structures, in quasi one-dimensional (1D)
and higher-dimensional settings respectively.
In that direction, and in connection with the above setting
of interfering condensates, a very interesting
observation was originally
reported in Ref.~\cite{reinhardt}. In particular,
it was numerically
found that
the collision of
two quasi-1D BEC fragments upon release of the light sheet
may lead to the formation of a train of dark solitons, filling the space
originally covered by the light sheet. The phase of the condensate
and its jumps around the soliton locations offered undisputed evidence
that this ``nonlinear interference'' leads to the formation of the
relevant localized nonlinear structures. Subsequent work in Ref.~\cite{ballagh}
aimed to clarify the regimes where the quasi-linear interference
of the original experiments \cite{science} would result, versus the ones where
the nonlinear interference of Ref.~\cite{reinhardt} forming dark soliton fringes,
would arise. Importantly, new relevant experiments were
very recently reported \cite{brian},
in which three independent BECs interfered while trapped,
giving rise to the formation of vortices.
The latter experiment can rather naturally be characterized as
a three-dimensional generalization of the original proposal
of  Ref.~\cite{reinhardt}.

In the present work, we revisit the nonlinear interference problem
of Ref.~\cite{reinhardt} in its
higher-dimensional version, relevant to
two-dimensional (2D) BECs (which are experimentally accessible \cite{pancake}),
as well to fully three-dimensional (3D) BECs. We use appropriately crafted potentials to slice the
condensate in two, as well as in {\it four} parts in 2D
settings and observe their nonlinear interference in these
cases. Then, we fragment the 3D BEC into {\it two, four} and
{\it eight} pieces and examine the results of their
merging as well.
The main finding of the present work is that
the nonlinear interference of higher-dimensional BECs
typically give rise to the higher-dimensional analogs
of the 1D train of dark solitons.
Specifically, in 2D, we find nucleation of
vortex-antivortex pairs and vortex necklaces
(which have previously been predicted to be formed in BECs
as a result of the snaking instability
of ring dark solitons \cite{RDS}), while in 3D
we find that the vortex patterns are replaced by vortex ring ones.
The patterns created by the collision of different
BEC slices become more complex for larger numbers of slices.
In examining the robustness of the relevant mechanism, we also
explore the role (in the vortex formation process)
of the relative phases of the different fragments,
as well as those of dissipation (emulating the interaction of the
condensate with a non-condensed atom fraction) and of the time for ramping
down the barrier separating the fragments.

Our presentation will be structured as follows.
First, in Sec.~\ref{2D}, we
present the 2D version of the problem:
the condensate is
separated into two fragments by a light sheet and then allowed
to collide (after light sheet removal), producing vortex
structures.
In Sec.~\ref{2D}, we also present
results pertaining to the
collision of four fragments separated by a light ``cross''.
Next, in Sec.~\ref{3D}, we systematically study the 3D case and interpret
the corresponding findings for, respectively, two, four and
eight condensate fragments.
Finally, in Sec.~IV, we
summarize and present our conclusions, as well as discuss some
interesting aspects meriting future study.

\section{Two-Dimensional Condensates\label{2D}}
\subsection{Collision of two in-phase fragments}

In the 2D case, we propose two different experimentally
feasible situations. In the more standard one, i.e., the direct analog
of the 1D case of Ref.~\cite{reinhardt}, the potential in Eq.~(\ref{ceq1}) reads
\begin{eqnarray}
V(x,y)= V_{\rm HT}(x,y) + V_{\rm LS}(x,y) ,
\label{ceq2}
\end{eqnarray}
where, for the 2D case,
\begin{eqnarray}
V_{\rm HT}(x,y) &=& \frac{1}{2}\, \Omega^2 (x^2+y^2) , \\[1.0ex]
V_{\rm LS}(x,y) &=& V_0\, {\rm sech}(by),
\label{MTSL}
\end{eqnarray}
where
$V_{\rm HT}$
represents the
harmonic trapping potential
(with $\Omega$ being the normalized trap strength), while
$V_{\rm LS}$ is a localized repulsive potential describing
the light sheet (with $V_0$ and
$b$ representing the normalized intensity
and inverse width of the laser beam).
Note that, similarly to the experiments of Ref.~\cite{brian}, the presence of
$V_{\rm HT}$ is necessary to guarantee nonlinear interference of the two condensate fractions separated
by the light sheet potential $V_{\rm LS}$ (the latter pushes the BEC atoms away from the vicinity of the line $y=0$).

\begin{figure*}[ht]
\hskip 0.0cm (a) \hskip 8.6cm (b) \\
\vspace{-0.3cm}
\begin{center}
\includegraphics[height=8.00cm]{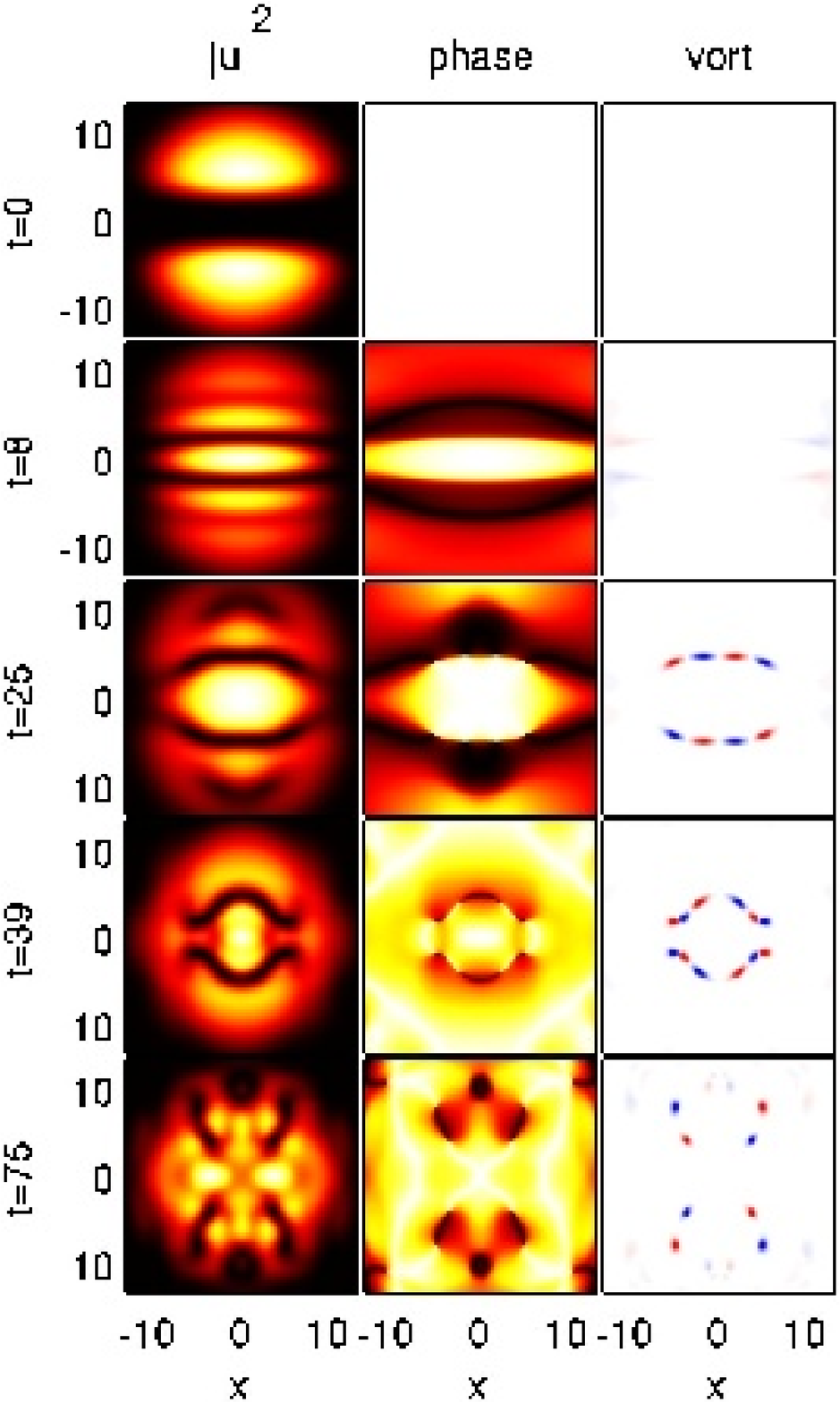}
\includegraphics[height=7.85cm]{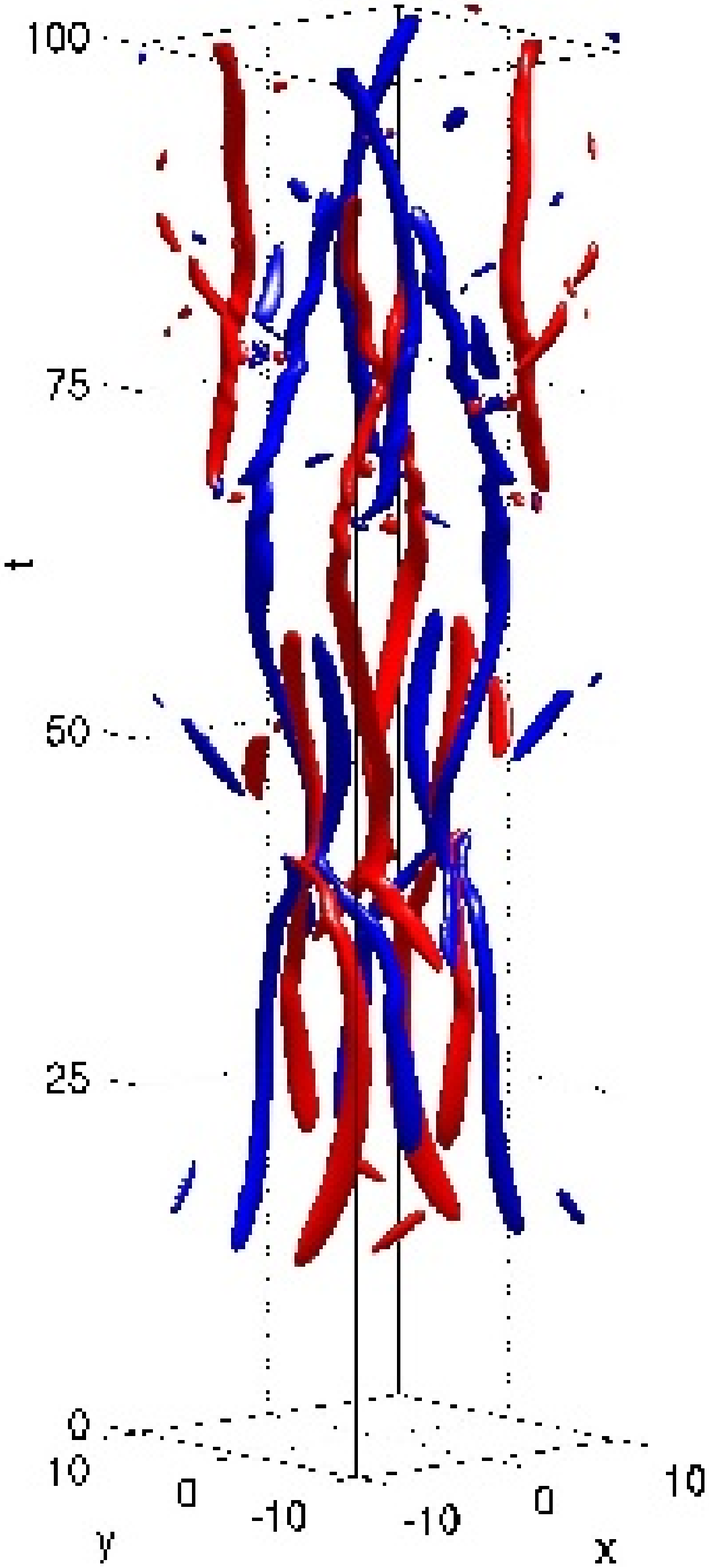}
~~~
\includegraphics[height=8.00cm]{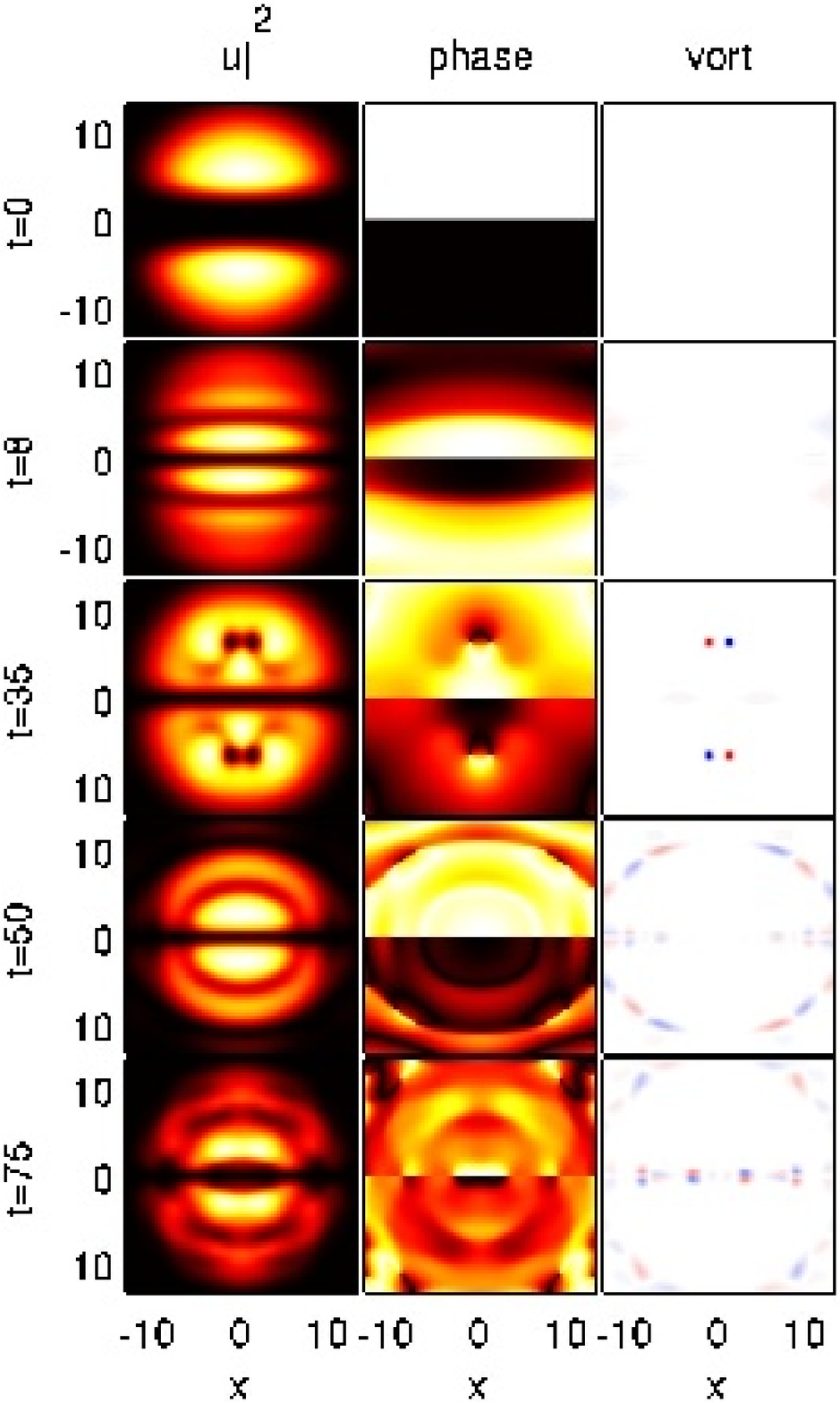}
\includegraphics[height=7.85cm]{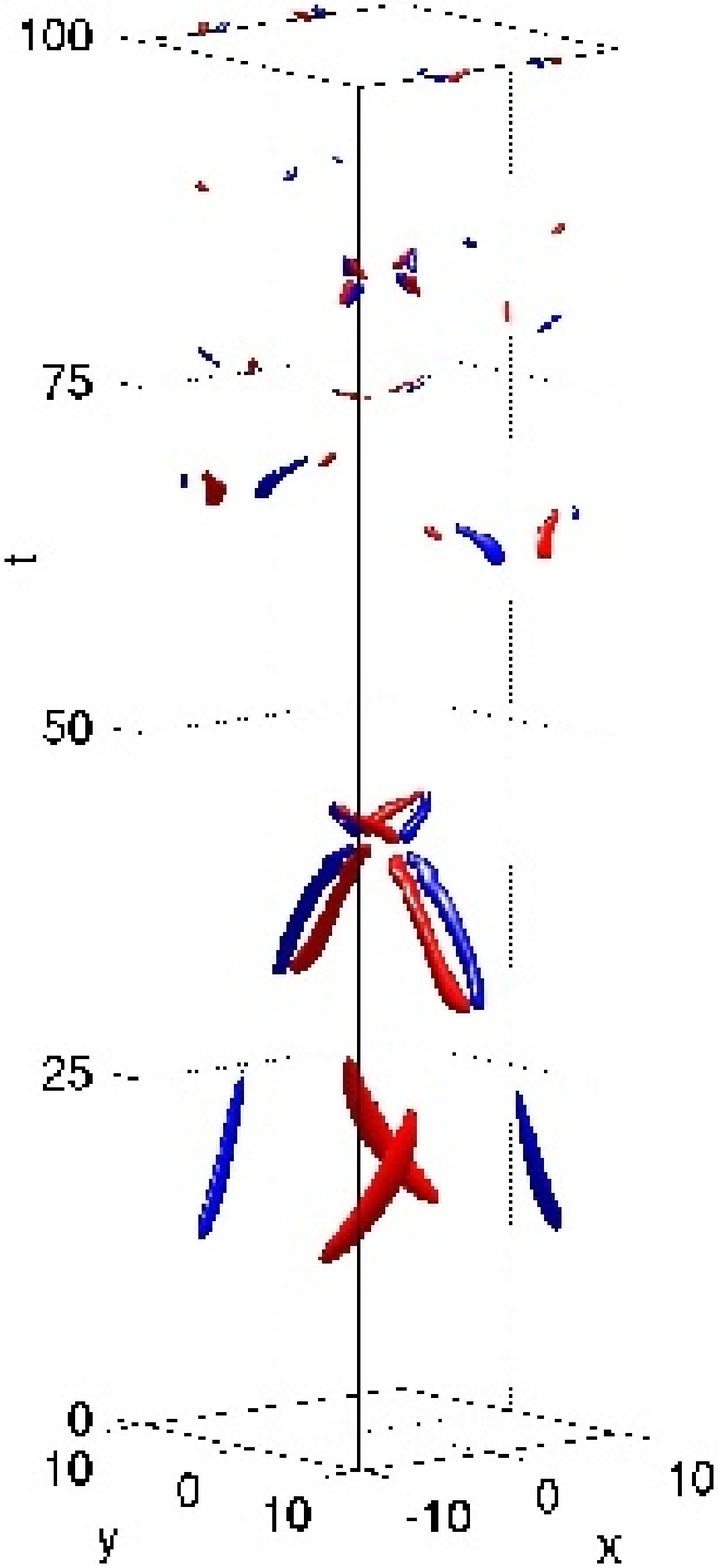}
\end{center}
\vspace{-0.3cm}
\caption{(Color online)
Collision of two condensate fragments originally separated
by a quasi-one-dimensional light sheet lifted at $t=0$.
Panels (a) correspond to fragments with same initial phase
($\Delta\phi=0$) while panels (b) correspond to
$\Delta\phi=\pi$.
For each case,
the first column of panels depicts the density
of the condensate $|u(x,y,t)|^2$ at the times indicated
(brighter regions indicate higher densities).
The result of the release of the light sheet is
the interference between the two fragments, see $|u(t\!=\!8)|^2$,
that eventually leads to the nucleation of vortex pairs.
The second column of panels depicts the corresponding
phases (brighter regions indicate phases close to $-\pi$
and $+\pi$ while dark regions correspond to zero phase).
The third column of panels depicts the corresponding fluid
vorticity (blue/red indicates positive/negative vorticity),
see text.
%
The fourth panel depicts the spatio-temporal evolution of the
vorticity by showing a contour slice at a tenth of the
maximum vorticity.
The evolution is responsible for the nucleation and
annihilation of vortex pairs resulting in the spatio-temporal
vortex filaments shown in the panel.
In all the 2D simulations, we use, for the spatial variable,
a discretization of $301\times301$ sites centered around
$(x,y)=(0,0)$. Also, the trap parameters for all 2D
simulations are: $\Omega=0.1$, $V_0=5$, and $b=1$.
}
\label{fig2d_1}
\end{figure*}

In
the case under consideration, the ground state of the system
has the form shown in the
top left panel in Fig.~\ref{fig2d_1}.a, namely two identical condensate fragments
separated by the light sheet.
This ground state is obtained by relaxation (imaginary time integration) starting from a
Thomas-Fermi cloud with density
%
\begin{eqnarray}
|u_{\rm TF}(x,y)|^2 =\max\{0,\mu-V(x,y)\},
\label{ceq3}
\end{eqnarray}
where $\mu$ is the chemical potential that will be set to
unity in what follows. Our proposed experiment assumes that
the condensate, confined in the
the potential of Eq.~(\ref{ceq2})
is in the ground state at $t=0$,
and that both fragments share the {\em same} phase (see below
for a description when fragments have deferent phases).
Then, at that time, $t=0$, we
``lift'' (i.e., switch-off) the light sheet
by setting $V_0 \rightarrow 0$, and subsequently
let the system evolve
according to the GP Eq.~(\ref{ceq1}). Notice that this is the direct
generalization in 2D
of the
setup proposed in Ref.~\cite{reinhardt}.
During the evolution, the two fragments originally constituting the condensate expand,
and eventually
interfere [see later times in
the left column of Fig.~\ref{fig2d_1}(a)].
In order to quantify the amount of vorticity generated by the
collision of the different fragments we monitor the vorticity
$\omega({\bf r},t)$ [$\bf r$ corresponds to $(x,y)$ in the 2D
simulations and $(x,y,z)$ in the 3D simulations]. The vorticity is
calculated as the curl
of the fluid velocity, ${\mathbf v}_s$,
namely $\omega({\bf r},t)=\nabla\times{\mathbf v}_s$
(see, e.g., \cite{Jackson:prl:98}), where the fluid velocity is given by:
\begin{equation}
{\mathbf v}_s=\frac{u^*\nabla u - u\nabla u^*}{i|u|^2}.
\label{fluid_vel}
\end{equation}
The typical numerical experiment showing the collision
between two fragments with same initial phase ($\Delta\phi=0$, see below) shown in
Fig.~\ref{fig2d_1}(a) is performed for $\Omega=0.1$, $V_0=5$ and $b=1$.
This choice corresponds, e.g., to a pancake sodium BEC, containing $N \approx 300$ atoms,
and confined in a trap with frequencies
$\Omega_r = 2 \pi \times 10$Hz and $\Omega_z = 2 \pi \times 100$Hz. On the other hand,
the light sheet parameters may correspond, e.g., to a blue-detuned laser beam providing a maximum
barrier energy of $k_B \times 24$~nK. Note that the phenomenology
that will be
presented below does not
change for the more realistic case of smaller trap strengths $\Omega$, which leads to larger numbers of atoms.

Through the interference, quasi-1D
``nonlinear'' fringes are formed, i.e., dark stripes (see atom density at $t=8$ in
the left panel of Fig.~\ref{fig2d_1}.a) resembling dark solitons (in direct analogy to the
1D case), but now in the 2D setting.
However, it is well-known that such 1D stripes are unstable in 2D (and 3D)
towards transverse modulations (see, e.g.,
Refs.~\cite{kivshar,shlyap,pra,mplbp} and references therein). As a result
of the ensuing snaking instability, such stripes break up into
vortex-antivortex pairs (because the original zero vorticity of the solution
needs to be preserved). This is precisely what happens in our
case as well. The bending of the stripes [see the atom
density at $t=25$ in Fig.~\ref{fig2d_1}(a)]
is responsible for the break up into four vortex pairs
as is shown in vorticity plot at $t=25$.
As the dark stripes continue bending and mixing, a series
of vortex pair nucleations and annihilations occur.

To follow this evolution, we depict in the right panel of Fig.~\ref{fig2d_1}(a)
a space-time plot of the vorticity. The main evolution of the
vorticity can be summarized as follows.
%
At $t\approx 15$ four vortices (i.e., two vortex pairs) nucleate. 
Two more quartets of
vortices are nucleated at $t\approx 22$ and $t\approx 35$.
At $t\approx 45$ the first quartet merges and is annihilated.
The second quartet merges at $t\approx 58$.
This process of merging and spontaneous nucleation of
vortex pairs and quartets remains active for long times (data not shown here).




\begin{figure*}[ht]
\hskip 0.0cm (a) \hskip 8.6cm (b) \\
\vspace{-0.3cm}
\begin{center}
\includegraphics[height=8.00cm]{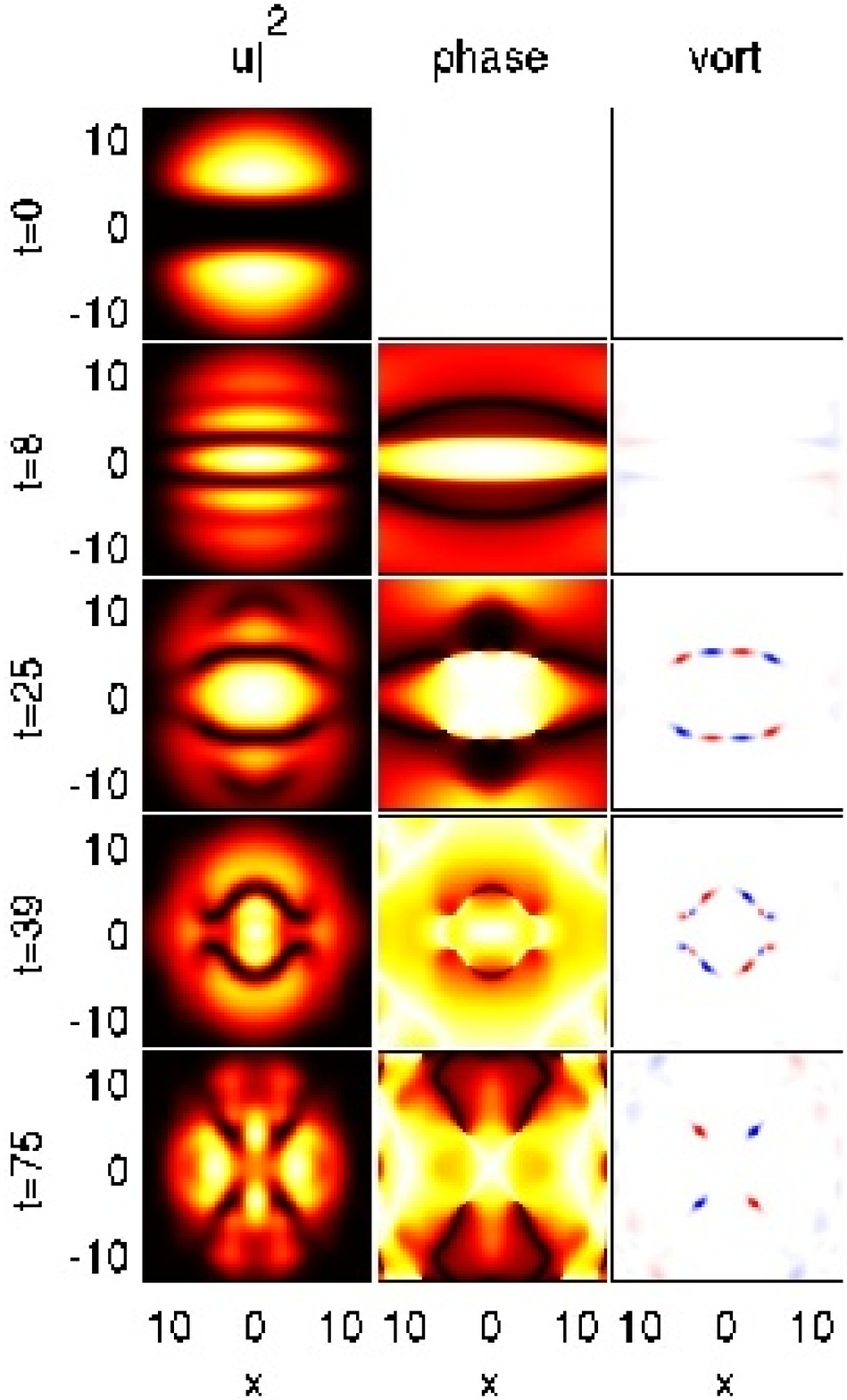}
\includegraphics[height=7.85cm]{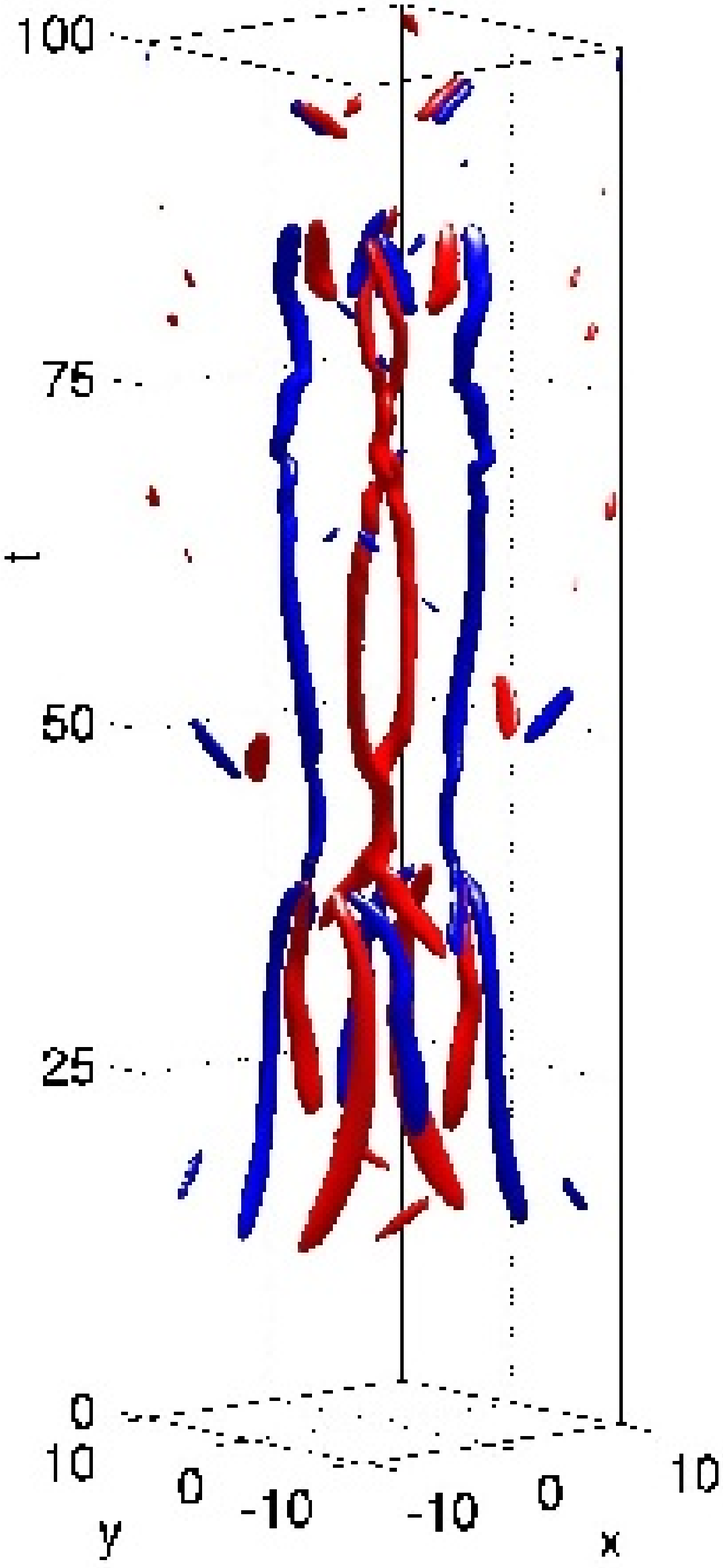}
~~~
\includegraphics[height=8.00cm]{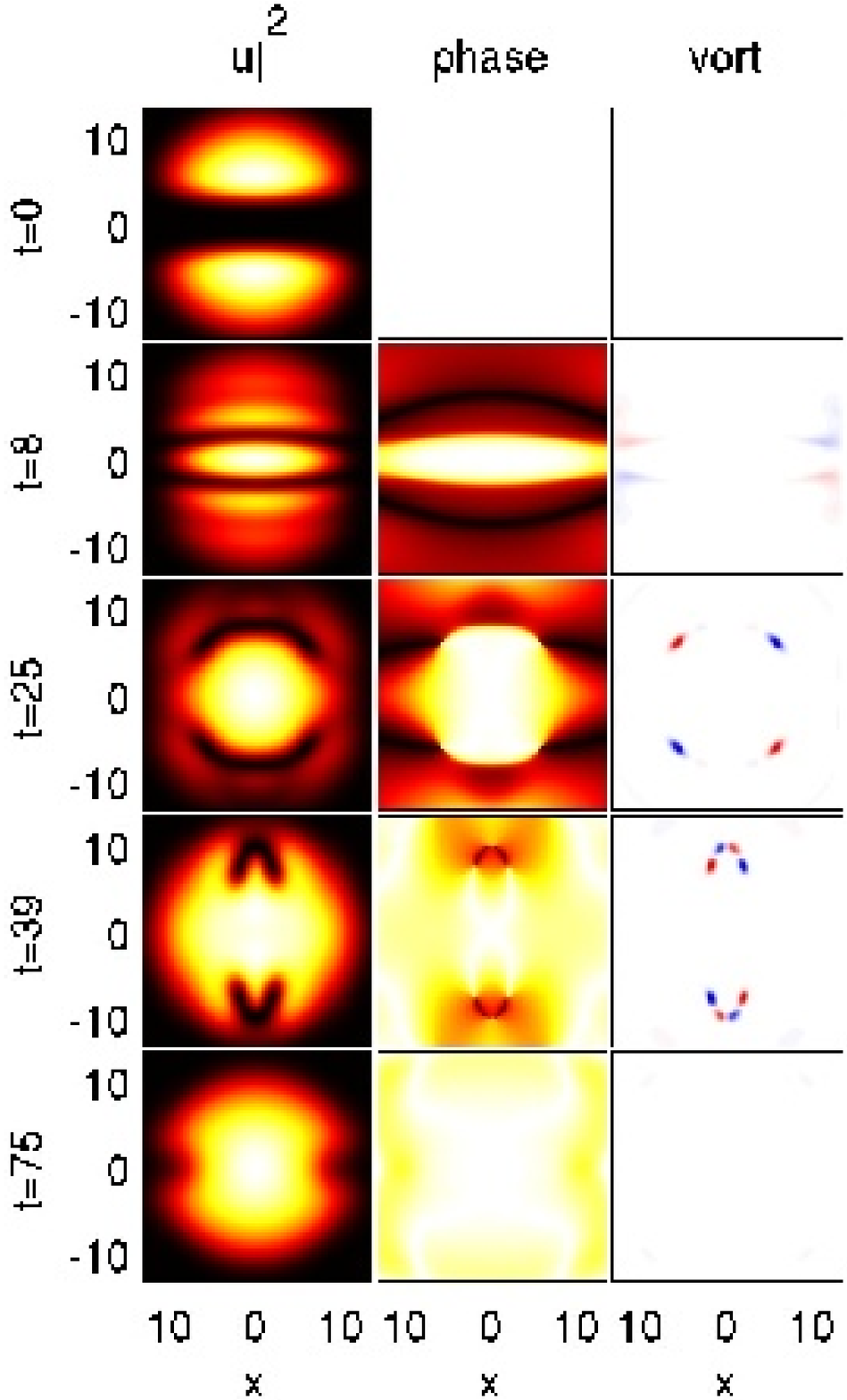}
\includegraphics[height=7.85cm]{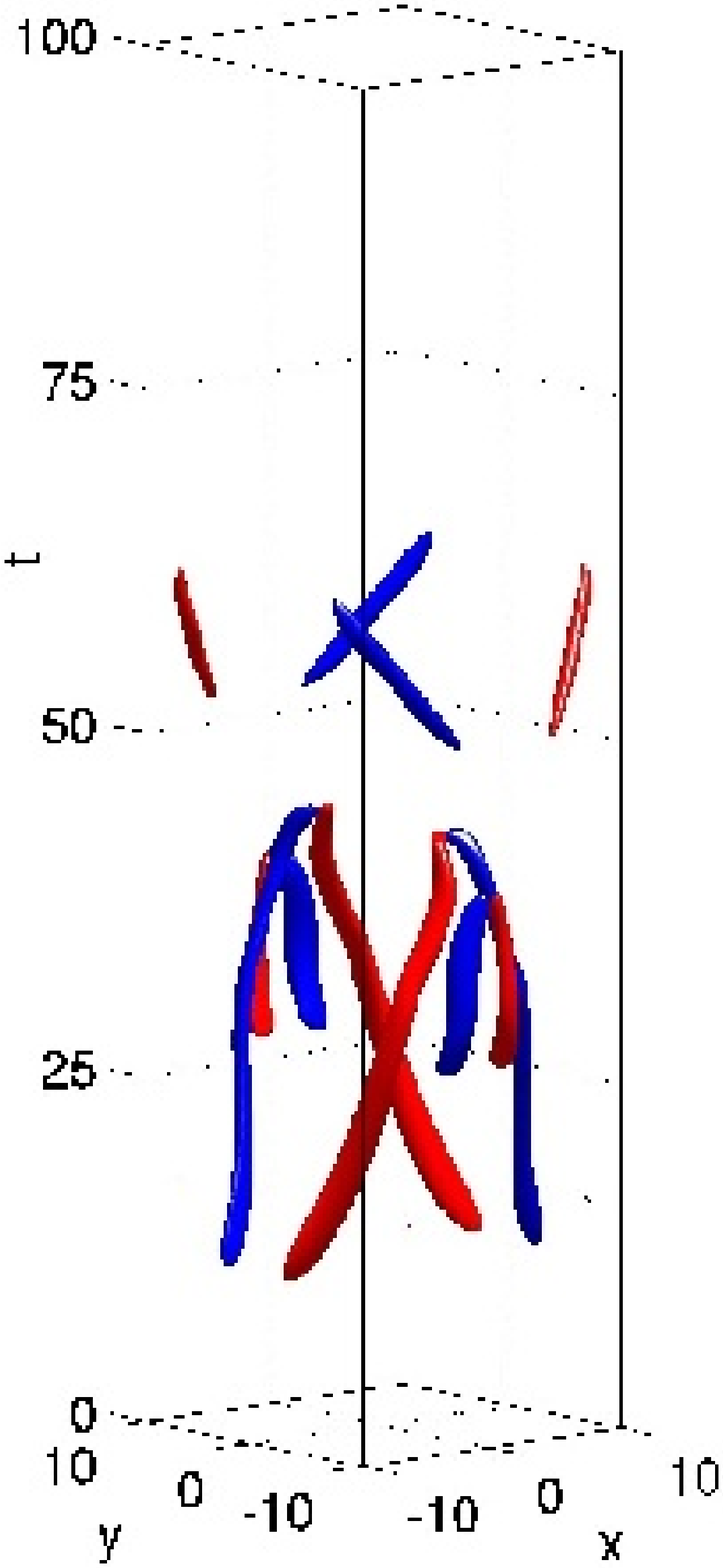}
\end{center}
\vspace{-0.3cm}
\caption{(Color online)
Effects of phenomenological damping on the
collision of two condensate fragments
with same initial phase ($\Delta\phi=0$).
The panels depict the same information as in Fig.~\ref{fig2d_1}(a)
with the addition of phenomenological damping
with (a) $\gamma=0.005$ and (b) $\gamma=0.1$.
}
\label{fig2d_damp}
\end{figure*}

\subsection{
The role of phase in two fragments collisions}

In actual BEC experiments (cf.~Ref.~\cite{brian}) the condensate
is grown inside the {\em combined} harmonic
trap $V_{\rm HT}$ and light sheet potential $V_{\rm LS}$.
There are two distinct regimes depending
on the strength of the light sheet: if the light sheet strength is
weak, the fragments have enough overlap and maintain a common phase,
however, for stronger light sheets, the fragments grow essentially
independently carrying their own independent (random) phases. Another
possibility in the experiment is to grow a condensate in the
harmonic
trap alone and then split the condensate into fragments
by adiabatically ramping up the light sheet. Depending on the degree
of adiabaticity and, more importantly, on the time the fragments
are kept separate before releasing the light sheet, the phases of
the different fragments will evolve independently and assume, effectively,
random phases. Here, we explore the effects of such phase differences
between the condensate fragments. We therefore consider the
case where just
before the light sheet release the phase of the
top (bottom) fragment is $\phi_1$ ($\phi_2$).
Without loss of generality  we fix $\phi_1=0$ and we focus on
the phase difference between the condensates $\Delta\phi\equiv\phi_2-\phi_1=\phi_2$.
In Fig.~\ref{fig2d_1}(b) we depict the collision of two
fragments as in Fig.~\ref{fig2d_1}(a) but with a different initial
relative phase between the condensates of $\Delta\phi=\pi$.
In
the simulations, initial configurations with arbitrary phase differences
$\Delta\phi$ were achieved by using the steady state solution found
for $\Delta\phi=0$ as before,
then applying a phase shift of
$\Delta\phi$ to the bottom fragment (i.e., for $y<0$), and running
again
the relaxation scheme (imaginary time integration) until
convergence.

The results presented in Figs.~\ref{fig2d_1}(a) and \ref{fig2d_1}(b)
indicate that
the detailed evolution of the vortex formation and merging depends
on the relative phases of the condensates.
In fact, further numerical results with phase differences between
$\Delta\phi=0$ and $\Delta\phi=\pi$ show similar vortex formation
and annihilation with vortex ``activity'' decreasing (both
in terms of the number of vortices produced and the persistence
time of the structures) as
$\Delta\phi$ was increased from 0 to $\pi$ (results not shown here).
It is interesting to note that the dynamics for phase differences
different from zero loses its four-fold symmetry when compared
to the $\Delta\phi=0$ case.
In all cases that we tried, vortex pairs are nucleated at some point
in the simulation irrespective of $\Delta\phi$.
However, it is evident from the figures that the smaller the phase
difference between the fragments, the more vortex structures are nucleated
and
the longer they live.
In fact, the extreme case of a $\Delta\phi=\pi$ between fragments
barely produces any vortex pairs [see Fig.~\ref{fig2d_1}(b)]. 
This is due to the fact that the
initial condition is close to a steady-state domain-wall (dark soliton
stripe separating out of phase domains)
and thus there is little extra energy for the collision of the
fragments. Indeed, as it can be evidenced from Fig.~\ref{fig2d_1}(b)
(where $\Delta\phi=\pi$),
the condensate always maintains a configuration similar to a
domain-wall with some perturbations: the density always shows
a nudge at $y=0$ and there is a predominant phase difference of
$\pi$ between the upper and lower half planes.

\subsection{Dissipation and barrier ramping-down effects}

The GP model used above relies, by construction, on the mean-field
description of a boson gas
at
extremely low temperatures and becomes exact at $T=0$.
When the temperature is finite, but still below the critical temperature
$T_c$ for BEC formation, there exists a fraction of
atoms that is not condensed, the so-called thermal cloud.
This thermal cloud is in fact coupled
to the condensed gas
and its presence
produces effects that
are not accounted for by the GP equation (cf.~the insightful review in
Chap.~18 of Ref.~\cite{BECBOOK} and references therein).
%
Phenomenologically, one of the most noticeable effects of the
presence of a sizeable thermal cloud is the
introduction of damping (to the condensed gas).
The approach of adding phenomenological damping to emulate thermal effects
was originally
proposed by Pitaevskii \cite{PitaevskiiPhenomenology} and applied with
a position-dependent loss rate in Ref.~\cite{ChoiPhenomenology}.
In practice, a few different implementations for the phenomenological
damping are viable.
Here, we follow the approach
of Refs.~\cite{Ueda02,Ueda03} where the GP equation is
modified by the inclusion of a damping term,
thus taking the form
\begin{equation}
( 1 - i \gamma) \frac{\partial}{\partial t} u
= -\frac{1}{2} \nabla^2 u + V({\bf r};t) u + |u|^2 u -\mu u,
\end{equation}
where $\gamma$ is the damping rate (the chemical potential is
again fixed at $\mu=1$).
In Fig.~\ref{fig2d_damp} we display the collision of two
fragments as in Fig.~\ref{fig2d_1}(a) (i.e., same initial phase,
$\Delta\phi=0$) for two values of damping. Figures~\ref{fig2d_damp}(a)
and \ref{fig2d_damp}(b) correspond to the cases of weak
($\gamma=0.005$) and strong ($\gamma=0.1$) damping.
For the above choices, it is relevant to
note that in Ref.~\cite{ChoiPhenomenology} it was found that
the value $\gamma=0.03$ corresponds to a temperature of about
$0.1T_c$.
As it can be observed from the figure, the case of weak damping
[$\gamma=0.005$, see Fig.~\ref{fig2d_damp}(a)]
behaves qualitatively the same as the
case without damping [see Fig.~\ref{fig2d_1}(a)] until about $t=40$,
when a noticeable reduction in the generated vorticity sets in.
This effect is even more dramatic for larger damping
[$\gamma=0.1$, see Fig.~\ref{fig2d_damp}(b)] where it can
be seen that after $t=65$ there is a complete absence
of vortices.
The effect of damping can be easily, qualitatively, understood in that
the added phenomenological dissipation slows the fragments in their
collisions and, therefore, less vorticity is produced.
%

We also studied the effect of ramp-down time of the laser sheet
separating the different condensate fragments. In current experiments
(see,
e.g, Ref.~\cite{brian}) the barrier created by the laser sheet can be
removed using a gradual ramp-down of the laser intensity.
Simulations (not show here) reveal that slower ramping-down
times have similar suppressing effects on vortex formation
to the phenomenological damping cases
described above. In fact, this is also for similar reasons, namely
that longer ramping-down times induce a slowing down of the
fragments and hence the production of less vorticity.
A more detailed study of the effects of
ramp-down times on the vortex nucleation is beyond the
scope of the present manuscript and will be reported in a separate
publication \cite{brian_plus_us}.

\begin{figure}[t]
\begin{center}
\includegraphics[height=8.00cm]{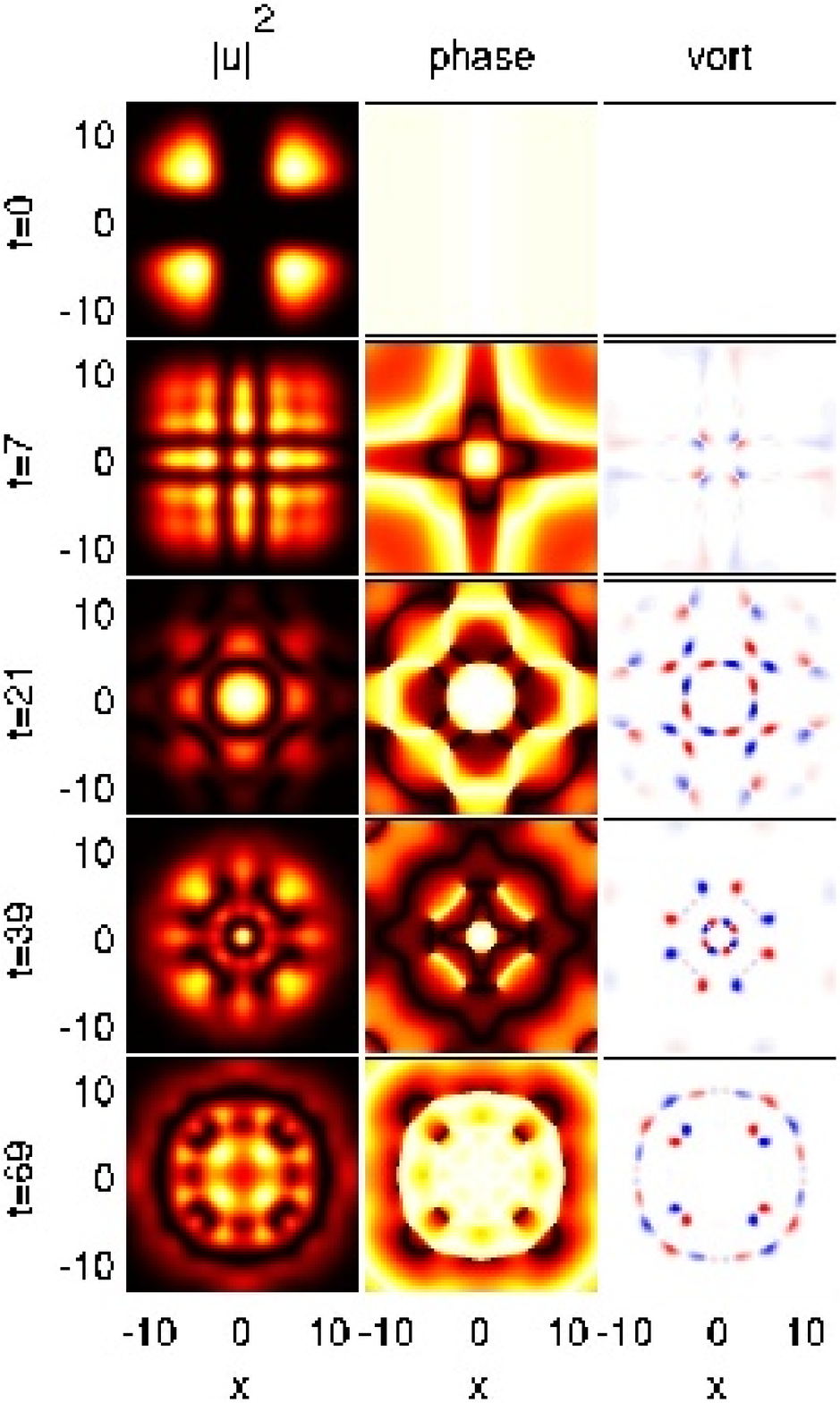}
\includegraphics[height=7.85cm]{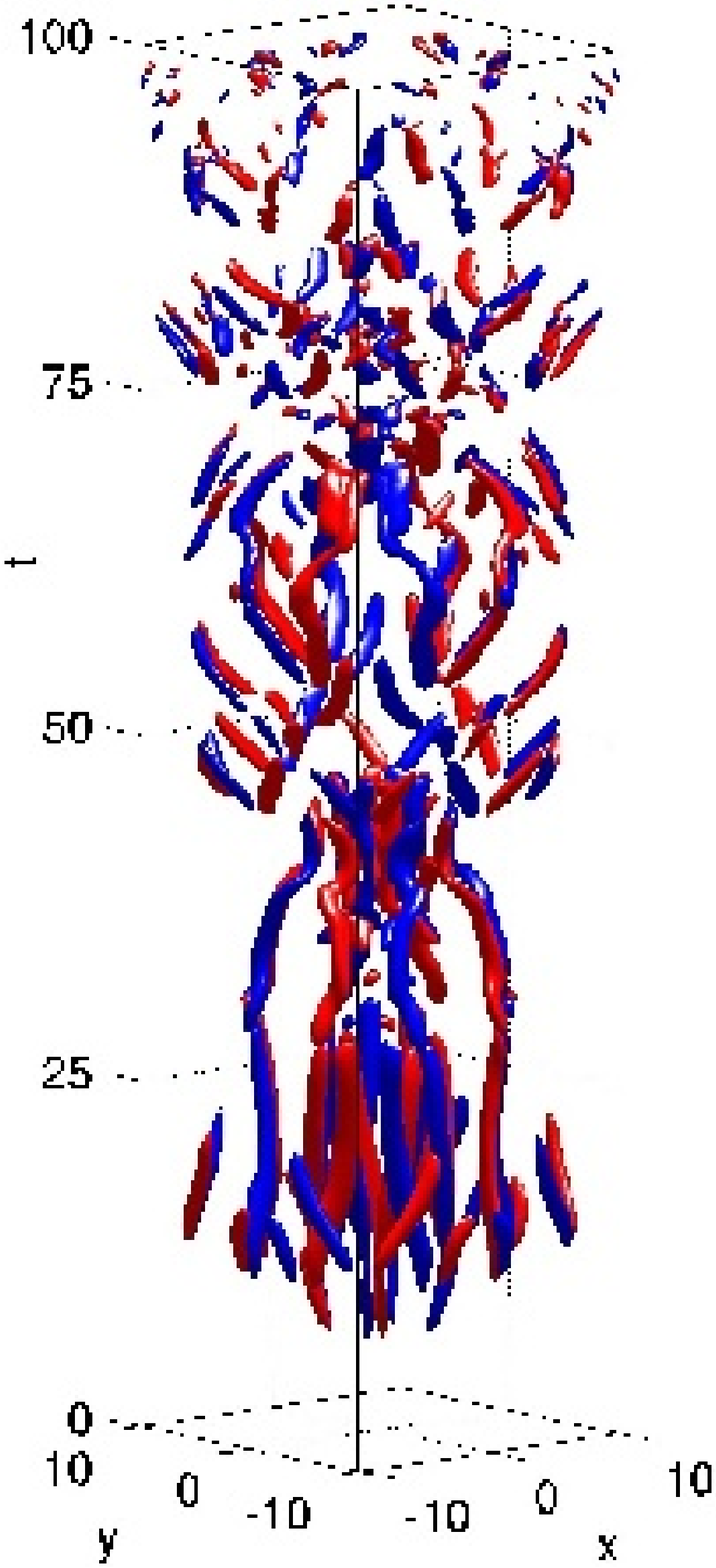}
\end{center}
\vskip-0.6cm
\caption{(Color online) Similar to Fig.~\ref{fig2d_1}(a), but for an
initial condition that contains four fragments (see $|u(t\!=\!0)|^2$),
being the ground state of the potential of Eq.~(\ref{ceq4}). Notice the four-fold
symmetry of the interference in this case and the eventual
formation of a complex pattern of vortex pairs.
Same parameters as in Fig.~\ref{fig2d_1} with $a=b=1$.
}
\label{fig2d_quad}
\end{figure}

\subsection{Collision of four in-phase fragments}

The above cases represent 2D renderings
of the setting as originally proposed in Ref.~\cite{reinhardt}
(which, however, possesses a number of
twists particular to 2D, as illustrated above).
Nevertheless, a more genuinely 2D
installment of the same ``experiment'' can also be envisioned and
is proposed in Fig.~\ref{fig2d_quad}. In this case, the light-sheet
potential is of the form
\begin{eqnarray}
V_{\rm LS} = V_0\, \left( {\rm sech}(a x) + {\rm sech} (b y) \right),
\label{ceq4}
\end{eqnarray}
resulting in a ground state which contains four in-phase condensate
fragments (as is shown by the initial density $|u(t\!=\!0)|^2$ in
Fig.~\ref{fig2d_quad}).
Subsequently, as the ``light cross'' is lifted at
$t=0$, the fragments attempt to fill in the empty space,
resulting in an interference pattern with four-fold symmetry.
The oscillation and ensuing bending of
the resulting dark stripes results in a rich evolution of
vortex pairs as it can be evidenced by the
space-time plot of the vorticity in the right panel of Fig.~\ref{fig2d_quad}.
Notice that for this initial condition, more exotic patterns,
including {\it vortex necklaces} (cf. also Ref.~\cite{RDS}) and structures
with a higher number of vortices are formed.
However, these patterns are considerably less long-lived
as compared to the ones observed
in the two in-phase fragments numerical experiment
described above. A more detailed analysis of the
role of different relative phases for the different fragments
will be reported elsewhere \cite{brian_plus_us}.

\section{Three-Dimensional Condensates\label{3D}}

\begin{figure}[tbp]
\begin{center}
\includegraphics[width=8.5cm]{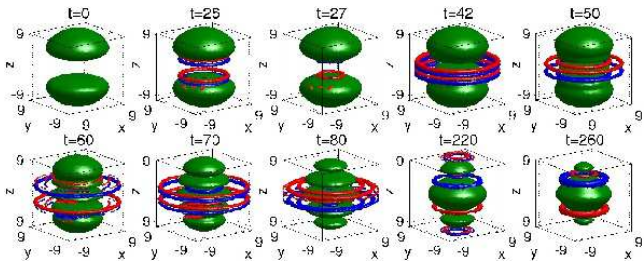}
\end{center}
\vskip-0.4cm
\caption{(Color online)
Collision of two 3D BEC fragments.
A 3D contour plot of atomic density $|u(x,y,z,t)|^2$
at $\max(|u|^2)/2.5$ and the vorticity
$\omega(x,y,z,t)$ at $\max(\omega)/15$
is shown as a function of $(x,y,z)$ at different times.
Positively (negatively) charged vortex rings are depicted
in blue (red).
%
The spatial discretization
is $50\times50\times50$ sites.
}
\label{fig3d_1}
\end{figure}

We now turn to the 3D analogs of the experiments proposed in the previous section.
We start, once again, by a 3D version of the 1D suggestion
of Ref.~\cite{reinhardt}, with a harmonic trapping potential
\begin{eqnarray}
V_{\rm HT} = \frac{1}{2}\, \Omega^2 (x^2+y^2+z^2),
\label{MT3D}
\end{eqnarray}
and a light sheet potential of the form
\begin{eqnarray}
V_{\rm LS} = V_0\, {\rm sech}(c z).
\label{ceq5}
\end{eqnarray}
We use $c=1$, in the typical results of Fig.~\ref{fig3d_1},
leading to the fragmentation of the initial condition into two
pieces as is shown for $t=0$. Once the single light sheet of
the potential of Eq.~(\ref{ceq5}) is lifted, interference ensues among the
different condensate fragments (see snapshots at $t=60,70,80$).
As the two fragments get closer, pairs of {\it vortex rings} are
nucleated between the fragments.
It is interesting to observe that these vortex rings,
created in pairs (recall that our system conserves
angular momentum) around the condensate (as smoke rings that are
created at the edge of the flow tube), may travel inward
and ``pinch'' the condensate promoting its fragmentation.
The formation of vortex rings in our system is reminiscent
of the experimental
realization of vortex rings induced by defects in
Ref.~\cite{hau}.
Here, again, a complex process involving nucleation and annihilation of
vortex rings persists for long times.

\begin{figure}[tbp]
\begin{center}
\includegraphics[width=8.5cm]{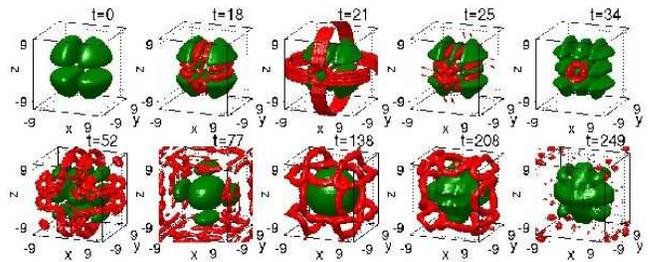}
\end{center}
\vskip-0.6cm
\caption{(Color online)
Similar to Fig.~\ref{fig3d_1}, but for an initial condensate
with four fragments. In this case, both, positively and
negatively charged vortex rings, are depicted in red.
Same parameters as in Fig.~\ref{fig3d_1} with $a=c=1$.
}
\label{fig3d_2}
\end{figure}

We now turn to the 3D analog of our second 2D numerical experiment,
using a ``light cross'' potential
\begin{eqnarray}
V_{\rm LS} = V_0\, \left( {\rm sech} (c z) + {\rm sech} (a x) \right).
\label{ceq6}
\end{eqnarray}
to split the 3D condensate
into four fragments initially, as shown at $t=0$ in Fig.~\ref{fig3d_2}.
These four fragments start interfering, producing  patterns
with fringes possessing four-fold symmetry (see snapshots
at $t=18,21,25,34$).
A similar scenario, though more complex, as in the previous experiment
takes place:
in particular, we observe the formation of vortex rings around
the condensate that promote the ``pinching'' of the different
interference fragments.
It is very interesting to note that for longer times ($t>100$)
a recurrent evolution emerges whereby a cage of vortex rings surrounds
the bulk of the condensate atoms (cf.~snapshots for
$t=138$ and $t=208$) alternated by
turbulent-like patterns with vortices (cf.~snapshots for $t=77$ and $t=249$).
This behavior seems to persist for long times (data not shown here).

\begin{figure}[tbp]
\begin{center}
\includegraphics[width=8.5cm]{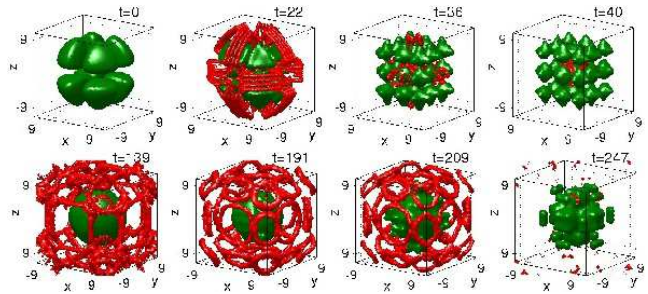}
\end{center}
\vskip-0.6cm
\caption{(Color online)
This last 3D case is similar to the earlier ones,
but for an initial condensate with eight fragments.
In this case also, both, positively and negatively charged vortex rings,
are depicted in red.
Same parameters as in Fig.~\ref{fig3d_1} with $a=b=c=1$.
}
\label{fig3d_3}
\end{figure}

Finally, in the same spirit as followed above, one can attempt
to produce a genuinely 3D version of this setup, by splitting
the condensate into 8 fragments as is shown in
Fig.~\ref{fig3d_3}. This can be done through a 3D light
sheet (consisting of three mutually perpendicular
1D light sheets):
\begin{eqnarray}
V_{\rm LS} = V_0\, \left( {\rm sech} (a x)+ {\rm sech} (b y) 
             + {\rm sech} (c z) \right).
\label{ceq7}
\end{eqnarray}
As a result of releasing the light sheets, the fragments interfere
with an eight-fold symmetry (see snapshots at $t=36,40$ in
Fig.~\ref{fig3d_3}). Again we find that vortex rings are created
in pairs around the condensate and subsequently
migrate inward promoting the ``pinching'' of the interference
pattern.
We also find the same recurrence phenomenon, as
in the previous experiment with four fragments, whereby a cage of vortex rings
surrounds the bulk of the condensate (cf.~snapshots at
$t=139,191,209$) alternating with vortex-rich, turbulent-like states
(cf.~snapshot at $t=247$).
Yet again this alternating behavior seems to persist
for longer times (data not shown here).
%
A general observation for the experiments with the two and three light sheets
(in comparison with the more straightforward realization with one
such sheet) is that vortex rings have a shorter lifetime and are harder
to detect due to the complicated nature of the
dynamics. This observation is,
in principle, also true for the 2D case (as can be
seen by comparing the two light sheet case, with that of a single
light sheet). Hence, perhaps the most robust and straightforward
configuration for observing the nonlinear interference dynamics
and the formation of vortices and vortex rings respectively in 2D
and 3D consists of the quasi-1D configuration with two fragments
(and a single light sheet)
in each case.

\section{Conclusions and Challenges}

In the present work, we have discussed a series of numerical
experiments, representing generalizations of the original suggestion
of Ref.~\cite{reinhardt} for a ``nonlinear interference'', leading to
the formation of fundamental nonlinear structures. Importantly,
relevant experiments with higher-dimensional
BECs (in a slightly different configuration than the one proposed here)
demonstrating this concept have been reported very recently \cite{brian}.

While in the 1D example of Ref.~\cite{reinhardt}, the ensuing
nonlinear waves were robust
dark solitons, in 2D and 3D settings the situation is
considerably more complex, even though fundamental nonlinear structures still
emerge. In the 2D setting, corresponding to a pancake condensate,
we find that slicing the condensate with a light sheet
or a light cross gives rise to interference patterns
and the nucleation of vortex pairs. This is the situation most reminiscent of
the experiment of Ref.~\cite{brian} (although the latter involved the
merger of three fragments), where vortex creation was observed as well.
In our case, subsequently, a complex cascade of vortex pair annihilations
and nucleations was found to persist for long times. Some vortex
``necklaces'' were also seen to be stable for long times.
This is particularly the case for a
BEC fragmented in two pieces, whereas the vortex patterns in the
four fragments case have shorter lifetimes, and
thus it is expected that it would be more difficult to observe
them experimentally.

In the simpler case of two fragments in 2D,
we studied the effects of different phases between
the condensate fragments.
An interesting extension of the results presented herein is the
study of the role of the phase difference between
fragments in a condensate with more than two fragments
\cite{brian_plus_us}.
We also analyzed the suppressing effects of phenomenological
damping (induced by coupling with the thermal cloud)
and briefly discussed the similar role of the ramping-down of
the light sheet on the formation of vorticity.

The 3D setting is particularly interesting, since the 2D vortex
patterns are replaced by vortex ring ones. We find that these vortex rings,
upon nucleating in pairs around the condensate,
migrate inward promoting the ``pinching''
of the different interference patterns. Subsequently the
condensate evolves under a complex pattern of
nucleation and annihilation of vortex rings.
More interesting, however, is the observation that for longer
times the condensate seems to alternate between a pattern
consisting of a vortex ring cage around the bulk
of the condensed atoms and a pattern of turbulent-like
vorticity.

An immediately interesting ramification of our
study would
be to compare/contrast the situation where the condensate fragments
originate from one component to that where they may originate from
different components \cite{two_comp}. Such interference experiments,
in part, already exist \cite{hall} and have interesting consequences
regarding the character of the interaction
of the two-components. It would be interesting to examine whether
such interactions can lead to (possibly anti-correlated) vortices
and vortex rings in the two-component condensate.
Such studies are currently in progress and will be reported elsewhere.

PGK and RCG acknowledge the support of NSF-DMS-0505663.
PGK also acknowledges support from NSF-CAREER and NSF-DMS-0619492.




\end{document}